\documentclass[smallextended,preprint]{svjour3}

\usepackage{graphicx,psfrag}
\usepackage{epsfig}
\usepackage{amssymb}
\usepackage{latexsym}

\journalname{\textbf{arXiv} e-print}

\begin{document}

\title{Mass of Neutron Star in SdS space-time}

\author{Vinayaraj O K \and
        V C Kuriakose }

\institute{Vinayaraj O K \at
              \email{vinayaraj@cusat.ac.in}
           \and
           V C Kuriakose \at
              \email{vck@cusat.ac.in} \\
              \emph{Department of Physics, Cochin University of Science and Technology, Kochi-682022,
              India}
}

\date{\today}

\maketitle

\begin{abstract}
In this work we present a modified TOV equation which incorporates
the cosmological constant with regard to the recent astronomical
observations that the Universe is in a phase of accelerated
expansion. Using this modified TOV equation we considered the
structure of a neutron star in SdS space-time and calculated maximum
mass limit for neutron stars.


\end{abstract}

\section{Introduction}
\label{intro} Neutron stars can be included in the class of compact
stars which are stars with high mass and small radius and their
density will be very high. So we need to take into account effects
from general relativity, like the curvature of space--time in the
study of neutron stars. Baade and Zwicky were the first to propose
the idea of neutron stars (Baade and Zwicky, 1934),
 pointing out that they have very high density and small
radius. They also made a very important suggestion that neutron
stars would be formed in supernova explosions. We know that there
are two forces acting on a star, one of them is gravitation and the
second one arises from the pressure. The opposing force of
gravitation will be thermal pressure for ordinary stars and
degeneracy pressure for white dwarfs and neutron stars. Thus we can
obtain two coupled differential equations for mass and pressure
which tell their change with the radius of the star.
Tolman-Oppenheimer-Volkoff (TOV) equation (Tolman, 1939; Volkoff and
Oppenheimer, 1939)
 derived from the Einstein's field equation gives the pressure gradient
 of neutron stars. Numerical model of a neutron star with a
relativistic mean field theory was given by Walecka (Walecka and
Chin, 1974; Walecka, 1975).

Our Universe can be represented by SdS metric, since the recent
astronomical observations show that the expansion of our universe is
accelerating (Guzzo et al., 2008; Perlmutter et al., 1999; Riess et
al., 1998; Weinberg, 1972).
Thus Schwarzschild-de Sitter metric described with a positive
cosmological constant and having two horizons is considered to be a
model of this universe. In such a case, the use of TOV equation,
derived from the Einstein's field equation in the case of
accelerating universe, will be more apt for the theoretical
constructions of stellar models. Thus we will get a modified TOV
equation for pressure, for calculating the maximum mass of a neutron
star. Anisotropic models for compact self gravitating objects
have been studied extensively (Corchero, 2001; Dev and Gleiser,
2003; Herrera and Santos, 1997; Ivanov, 2002; Mak and Harko, 2003).
Our aim is to establish a modified TOV equation which is the prime
structure equation for compact stars.

In \S2 we will show the derivation of the new equation to replace
the TOV equation, starting with the standard metric for a
spherically symmetric star and the consecutive use of Einstein's
field equation in the case of accelerating universe. Then we will
show the numerical calculation of maximum mass of Neutron stars by
plotting the mass-radius relation.
\section{Derivation of the modified TOV equation}
\label{sec:1} We have the standard metric (Weinberg, 1972)
for a spherically symmetric star,
\begin{equation}
ds^{2}=-U(r)dt^{2}+V(r)dr^{2}+r^{2}d\theta ^{2}+r^{2}\sin ^{2}\theta
d\phi ^{2},  \label{1}
\end{equation}
where,
\begin{equation}
V(r)=\frac{1}{U(r)}.  \label{2}
\end{equation}

Ricci tensor is given by,
\begin{equation}
R_{\mu k}=\frac{\partial \Gamma _{\mu \lambda }^{\lambda }}{\partial
x^{k}}-\frac{\partial \Gamma _{\mu k}^{\lambda }}{\partial
x^{\lambda }}+\Gamma _{\mu \lambda }^{\eta }\Gamma _{k\eta
}^{\lambda }-\Gamma _{\mu k}^{\eta }\Gamma _{\lambda \eta }^{\lambda
}.  \label{3}
\end{equation}
Substituting for the Christoffel symbols, we will get the equations
as,
\begin{equation}
R_{rr}=-\frac{U^{\prime \prime }}{2U}-\frac{U^{\prime }}{4U}\left(
\frac{V^{\prime }}{V}+\frac{U^{\prime }}{U}\right)
-\frac{1}{r}\frac{V^{\prime }}{V},  \label{4}
\end{equation}
\begin{equation}
R_{\theta \theta }=-1+\frac{r}{2V}\left( -\frac{V^{\prime
}}{V}+\frac{U^{\prime }}{U}\right) +\frac{1}{V},  \label{5}
\end{equation}
\begin{equation}
R_{\phi \phi }\approx R_{\theta \theta },  \label{6}
\end{equation}
\begin{equation}
R_{tt}=-\frac{U^{\prime \prime }}{2V}+\frac{U^{\prime }}{4V}\left(
\frac{V^{\prime }}{V}+\frac{U^{\prime }}{U}\right) -\frac{U^{\prime
}}{rV}. \label{7}
\end{equation}
The result $R_{\phi \phi }\approx R_{\theta \theta }$ is the
consequence of the rotational invariance of the metric.

Einstein's field equation in the case of accelerating universe
(Weinberg, 1972)
is given by,
\begin{equation}
G_{\mu \nu }=-8\pi GT_{\mu \nu }+\Lambda g_{\mu \nu },  \label{8}
\end{equation}
where
\begin{equation}
G_{\mu \nu }=R_{\mu \nu }-\frac{1}{2}g_{\mu \nu }R.  \label{9}
\end{equation}
Substituting Eq. (\ref{9}) in Eq. (\ref{8}), we get
\begin{equation}
R_{\mu \nu }=-8\pi G\left({T}_{\mu \nu}-\frac{1}{2}g_{\mu \nu
}{T}_{\lambda }^{\lambda }-\frac{\Lambda }{8\pi G}g_{\mu \nu
}\right).   \label{10}
\end{equation}
From hydrodynamics, we have the relation for energy-momentum tensor
in the case of models with cosmological constant,
\begin{equation}
\tilde{T}_{\mu \nu }=\tilde{p}g_{\mu \nu }+\left(
\tilde{p}+\tilde{\epsilon }\right) u_{\mu }u_{\nu }, \label{11}
\end{equation}
where $\tilde{p}$-the modified pressure, $\tilde{\epsilon } $-the
modified total energy density and $u_{\mu }$-the velocity four
vector. Here
\begin{equation}
\tilde{p}=p-\frac{\Lambda }{8\pi G} \ \ \ \ \ \ \ \ \tilde{\epsilon
}=\epsilon +\frac{\Lambda }{8\pi G},  \label{12}
\end{equation}

Also the velocity vector is defined so that,
\begin{equation}
g^{\mu \nu }u_{\mu }u_{\nu }=-1.  \label{13}
\end{equation}%
Assuming fluid at rest, we have
\begin{equation}
u_{r}=u_{\theta }=u_{\phi }=0,  \label{14}
\end{equation}
\begin{equation}
u_{t}=-(-g^{tt})^{-\frac{1}{2}}=-\sqrt{U(r)}.  \label{15}
\end{equation}

Our assumptions of time independence and spherical symmetry imply
that $p$ and $\epsilon $ are functions only of the radical
co-ordinate $r$.

Using Eq. (\ref{10}), we will get
\begin{eqnarray}
R_{rr}=-\frac{U^{\prime \prime }}{2U}-\frac{U^{\prime }}{4U}\left(
\frac{V^{\prime }}{V}+\frac{U^{\prime }}{U}\right)
-\frac{1}{r}\frac{V^{\prime }}{V} \nonumber \\ =-4\pi G(\epsilon
-p)V-\Lambda V, \label{16}
\end{eqnarray}
\begin{eqnarray}
R_{\theta \theta }=-1+\frac{r}{2V}\left( -\frac{V^{\prime
}}{V}+\frac{U^{\prime }}{U}\right) +\frac{1}{V} \nonumber \\ =-4\pi
G(\epsilon -p)r^{2}-\Lambda r^{2}, \label{17}
\end{eqnarray}
\begin{eqnarray}
R_{tt}=-\frac{U^{\prime \prime }}{2V}+\frac{U^{\prime }}{4V}\left(
\frac{V^{\prime }}{V}+\frac{U^{\prime }}{U}\right) -\frac{U^{\prime
}}{rV} \nonumber \\ =-4\pi G(\epsilon +3p)U+\Lambda U,  \label{18}
\end{eqnarray}
Here single prime denotes first derivative and double prime denotes
second derivative.

We have the equation for hydrostatic equilibrium,
\begin{equation}
\frac{U^{\prime }}{U}=\frac{-2p^{\prime }}{p+\epsilon }.  \label{19}
\end{equation}

Let us derive the equation for V(r) alone. Using Eq. (\ref{16}), Eq.
(\ref{17}) and Eq. (\ref{18}), we will get
\begin{equation}
\frac{-V^{\prime }}{rV^{2}}-\frac{1}{r^{2}}+\frac{1}{Vr^{2}}=-8\pi
G\epsilon -\Lambda,   \label{20}
\end{equation}
which can be written as
\begin{equation}
\left( \frac{r}{V}\right) ^{\prime }=1-8\pi G\epsilon r^{2}-\Lambda
r^{2}. \label{21}
\end{equation}

Integrating,
\begin{equation}
V(r)=\left[ 1-\frac{2GM(r)}{r}-\frac{\Lambda }{3}r^{2}\right] ^{-1},
\label{22}
\end{equation}
where
\begin{equation}
M(r)=\int\limits_{0}^{r}4\pi r^{\shortmid 2}\epsilon (r^{\shortmid
})dr^{\shortmid }.  \label{23}
\end{equation}
Eq. (\ref{22}) tells that outside the star, the space-time looks
like SdS.

 Now, we can use Eq. (\ref{19}), Eq. (\ref{22}) and Eq.
(\ref{23}) to eliminate U(r) and V(r) from Eq. (\ref{17}), which
becomes

\begin{equation}
-r^{2}p^{\prime }=GM\left( r\right) \epsilon \left( r\right)
\frac{\left[ 1+\frac{p\left( r\right) }{\epsilon \left( r\right)
}\right] \left[ 1+\frac{4\pi r^{3}p\left( r\right) }{M\left(
r\right) }+\frac{\Lambda r^{3}}{GM\left( r\right) }\right] }{\left[
1-\frac{2GM\left( r\right) }{r}-\frac{\Lambda r^{2}}{3}\right]}.
\label{26}
\end{equation}

Thus the modified TOV equation is,
\begin{equation}
\frac{dp}{dr}=-\frac{GM\left( r\right) \epsilon \left( r\right)
}{r^{2}}\frac{\left[ 1+\frac{p\left( r\right) }{\epsilon \left(
r\right) }\right]\left[ 1+\frac{4\pi r^{3}p\left( r\right) }{M\left(
r\right) }+\frac{\Lambda r^{3}}{GM\left( r\right) }\right] }{\left[
1-\frac{2GM\left( r\right) }{r}-\frac{\Lambda r^{2}}{3}\right] }.
\label{27}
\end{equation}

This equation is different from TOV equation and this becomes TOV
equation for $\Lambda =0$.

Now we can use this new equation to find the maximum mass of neutron
stars. In this work, neutron stars are considered to be composed of
a Fermi gas of degenerate neutrons. The mass-radius and
density-radius relations are plotted and maximum mass limit is
checked.

We have the standard structure equations for a spherically symmetric
star which describe how the mass and the pressure of a star change
with radius:

\begin{equation}
\frac{dm}{dr}=\frac{4\pi r^{2}\epsilon (r)}{c^{2}},  \label{28}
\end{equation}

\begin{equation}
\frac{dp}{dr}=-\frac{G\epsilon (r)m(r)}{c^{2}r^{2}}, \label{29}
\end{equation}

where $G$\ is Newton's Gravitational constant, $\epsilon $\ is the
mass density, $m$\ is the mass up to radius $r$\ and $c$\ is the
velocity of light.

The modified TOV equation is similar to the differential equation
for the pressure, Eq. (\ref{29}), but here has three correction
factors. All three correction factors are larger than one, i.e. they
strengthen the term from Newtonian gravity. Here one has to solve
(Chandrasekhar, 1931)
the coupled differential Equations (\ref{28}) and (\ref{27}). We
must notice that while $\frac{dm}{dr}$\ is positive,
$\frac{dp}{dr}$\ must always be negative. Starting with certain
positive values for $m$\ and $p$\ in a small central region of the
star the mass will increase while the pressure decreases eventually
reaching zero. We can set $m(r=0)=0$\ and , in addition we specify
the central pressure $p(r=0)=p_{0}$\ in order to solve for Equations
(\ref{28}) and (\ref{27}). The behaviour of $m$\ and $p$\ as a
function of radius will become important for numerical calculations.

Both the coupled differential equations depend on energy density. So
we need a relation between pressure $p$ and energy density
$\epsilon$ which is known as the Equation of State(EoS) (Shapiro and
Teukolsky, 1983) 
for the neutron stars. For mass calculation, we will use the well
known EoS:
\begin{equation}
p=K\epsilon^\gamma,  \label{32}
\end{equation}
where $K$ is a constant and $\gamma$ is the polytropic exponent.
$\gamma$=$\frac{4}{3}$ in general relativistic case (Silbar and
Reddy, 2004).

We will give the value of cosmological constant (Carmeli and
Kuzmenko, 2001; Carroll, 2000) 
as $\Lambda$$\approx$$10^{-46}$$km^{-2}$.
\section{Result and Discussion}
\label{res}
 The mass-radius and density-radius relations obtained is
shown in Fig. \ref{1} and Fig. \ref{2}. For calculational purpose,
we put $G$=1 and $c$=1 and then pick the units of all dimensionful
quantities to be powers of kilometers.

Converting the results back from natural units, we will get the
maximum mass of neutron star $M$=0.74$M_{\odot}$ and radius
$R$=6.98$km$.

Comparing with the result that obtained when calculated using the
normal TOV equation, we found that the obtained values are lesser.
So we can conclude that there is a slight influence of cosmological
constant in the stellar structure. The influence of cosmological
constant will be higher if the value is higher than the recently
accepted one. Future advancements in dark energy research may
strengthen the importance of cosmological constant. The modified TOV
equation could also be used in the calculations of surface tension
of compact stars (Bagchi, et al.,
2005; Dey, et al., 1998).

\begin{figure}
  \includegraphics{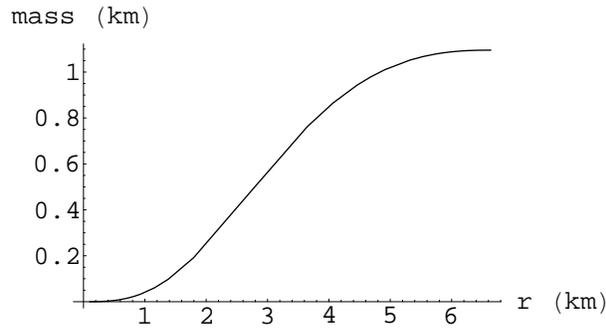}
\caption{Mass-Radius plot for neutron stars for a central density
$1\times 10^{16}gm/cm^{3}$.}
\label{fig:1}       
\end{figure}

\begin{figure}
  \includegraphics{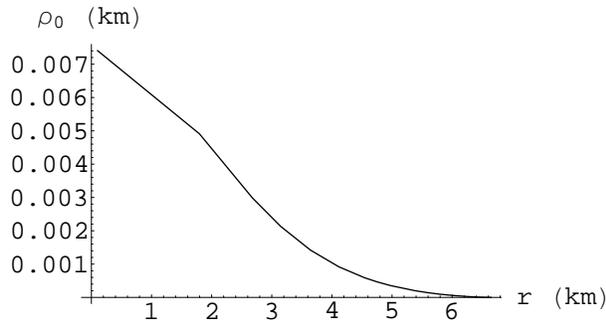}
\caption{Density-Radius plot for neutron stars for a central density
$1\times 10^{16}gm/cm^{3}$.}
\label{fig:2}       
\end{figure}


\begin{thebibliography}{}
Baade, W., Zwicky, F.: On Super-Novae. Proc. N. A. S., 254-259
(1934) \\
Bagchi, M., Sinha, M., Dey, M., Dey, J., Bhowmick, S.: Newtonian and
general relativistic contribution of gravity to surface tension of
strange stars. Astron. \& Astrophys. \textbf{440}, L33 (2005) \\
Carmeli, M., Kuzmenko, T.: Value of the Cosmological Constant:
Theory versus Experiment. astro-ph/0102033 \\
Carroll, S.M.: The Cosmological Constant. astro-ph/0004075 \\
Chandrasekhar, S.: The Maximum Mass of Ideal White Dwarfs.
Astrophys. J. \textbf{74}, 81 (1931) \\
Corchero, E.S.: Quantum Approach to Neutron Stars Leading to
Configurations With Local Anisotropy and Mass Above the
Oppenheimer-Volkoff Limit. Astrophysics \& Space Science
\textbf{275}, 259 (2001) \\
Dev, K., Gleiser, M.: Anisotropic Stars II: Stability. Gen. Relat.
Grav. \textbf{35}, 1435 (2003) \\
Dey, M., Bombaci, I., Dey, J., Ray, S., Samantha, B.C.: Strange
stars with realistic quark vector interaction and phenomenological
density-dependent scalar potential. Phys. Lett. B \textbf{438}, 123
(1998) \\
Glendenning, N.K.: Compact Stars: Nuclear Physics, Particle Physics
and General Relativity. Springer-Verlag, New York (1997) \\
Guzzo, L.: A test of the nature of cosmic acceleration using galaxy
redshift distortions. Nature \textbf{451}, 541 (2008) \\
Herrera, L., Santos, N.O.: Local anisotropy in self-gravitating
systems. Phys. Rep. \textbf{286}, 53 (1997) \\
Ivanov, B.V.: Maximum bounds on the surface redshift of anisotropic stars.
 Phys. Rev. D \textbf{65}, 104011 (2002) \\
Mak, M.K., Harko, T.: Anisotropic stars in general relativity. Proc.
Roy. Soc. Lond. A \textbf{459}, 393 (2003) \\
Perlmutter, S. et al.: Measurements of omega and lambda from 42
high-redshift supernovae. Astrophys. J. \textbf517, 565 (1999) \\
Riess, A. G. et al.: Observational evidence from supernovae for an
accelerating universe and a cosmological constant. Astron. J.
\textbf116, 1009 (1998) \\
Shapiro, S.L., Teukolsky, S.A.: Black Holes, White Dwarfs and
Neutron Stars: The Physics of Compact Objects. John Wiley \& Sons,
New York (1983) \\
Silbar, R.R., Reddy, S.A.: Neutron stars for undergraduates. Am. J.
Phys. \textbf{72}, 892 (2004) \\
Tolman, R.C.: Static Solutions of Einstein's Field Equations for Spheres of Fluid.
 Phys. Rev. \textbf{55}, 364 (1939) \\
Volkoff, G.M., Oppenheimer, J.R.: On Massive Neutron Cores. Phys.
Rev. \textbf{55}, 374 (1939) \\
Walecka, J.D., Chin, S.A.: An equation of state for nuclear and higher-density matter based on relativistic mean-field theory.
 Phys. Lett. B \textbf{52}, 24 (1974) \\
Walecka, J.D.: Equation of state for neutron matter at finite T in a relativistic mean-field theory.
 Phys. Lett. B \textbf{59}, 109 (1975) \\
Weinberg, S.: Gravitation and Cosmology: Principles and Applications
of the General Theory of Relativity. John Wiley \& Sons, New York
(1972) \\

\end{thebibliography}
\end{document}